\documentclass[aps,prl,twocolumn]{revtex4}

\usepackage{epsfig}

\usepackage{amsmath,amssymb}
\usepackage{color}

\def\CC{{\rm\kern.24em \vrule width.04em height1.46ex depth-.07ex
\kern-.30em C}}
\def\P{{\rm I\kern-.25em P}}
\def\NN{{\rm I\kern-.25em N}}
\def\RR{{\rm
         \vrule width.04em height1.58ex depth-.0ex
         \kern-.04em R}}
\def\id{{\rm 1\kern-.22em l}}
\def\ZZ{{\sf Z\kern-.44em Z}}

\newtheorem{pdef}{Definition}[section]

\newenvironment{eqblock}[2]{\beq\label{#2}\begin{array}{#1}}{\end{array}
                                \eeq}
\newenvironment{neqblock}[1]{\[\begin{array}{#1}}{\end{array}\]}
\newcommand{\braket}[2]{\langle #1 | #2 \rangle}
\newcommand{\ketbra}[1]{\ensuremath{| #1 \rangle \langle #1 |}}

\newcommand{\beqb}{\begin{eqblock}}
\newcommand{\eeqb}{\end{eqblock}} 
\newcommand{\nbeqb}{\begin{neqblock}}
\newcommand{\neeqb}{\end{neqblock}} 

\newcommand{\beq}{\begin{equation}}
\newcommand{\beqa}{\begin{eqnarray}}
\newcommand{\eeq}{\end{equation}}
\newcommand{\eeqa}{\end{eqnarray}}
\newcommand{\nbeqa}{\begin{eqnarray*}}
\newcommand{\neeqa}{\end{eqnarray*}}

\newcommand{\ket}[1]{| #1 \rangle}

\def\DJo{$\;$\kern-.4em \hbox{D\kern-.8em\raise.15ex\hbox{--}\kern.35em okovi\'c}}

\begin{document}

\title{The nine ways of four-qubit entanglement and their threetangle}
\author{Andreas Osterloh}
\affiliation{Institut f\"ur Theoretische Physik, 
         Universit\"at Duisburg-Essen, D-47048 Duisburg, Germany.}
\email{andreas.osterloh@uni-due.de}
\begin{abstract}
I calculate the mixed threetangle $\tau_3[\rho]$ for the reduced density matrices of 
the four-qubit representant states found in Phys. Rev. A {\bf 65}, 052112 (2002).
In most of the cases, the convex roof is obtained, except for one class,
where I provide with a new upper bound, which is assumed to be very close to the convex roof. 
I compare with results published in Phys. Rev. Lett. {\bf 113}, 110501 (2014).
Since the method applied there usually results in higher values for the upper bound, 
in certain cases it can be understood that the convex roof is obtained exactly, namely when the 
zero-polytope where $\tau_3$ vanishes shrinks to a single point. 
\end{abstract}

\maketitle

\section{Introduction}

Entanglement has become a central part of modern physics and hence, also quantifying this physical ressource 
has gained much relevance.
For two qubits, only one class of entanglement exists, and its entanglement could also be 
determined exactly using the concurrence\cite{Hill97,Wootters98}. Every entanglement measure
that is extended from pure to mixed states by means of the convex roof construction\cite{Uhlmann98} 
can be written as a function of this measure. 
A bit later, the threetangle\cite{Coffman00} was extracted as the residual tangle of a monogamy relation
for pure three qubit states\cite{Coffman00,Osborne06}. 
This was the first entanglement measure that could distinguish in a sharp way 
the two different ways three qubits can be entangled\cite{Duer00}.
This is a task which is only achievable by an $SL$-invariant entanglement measure, instead of incorporating the 
minimal unitary symmetry of entanglement\cite{MONOTONES}.
Examples for the latter are entanglement measures connected to the partial transpose criterion\cite{Peres96,Horodecki96,Vidal02,Jungnitsch11}.
It was only later that the relevance of the invariance with respect to the group $SL$ 
became clearer to the community\cite{Duer00,
VerstraeteDMV02,VerstraeteDM02,VerstraeteDM03,OS04,OS05,DoOs08,GourWallach10}.
Unfortunately, the convex roof extension to mixed states seems difficult in view of the infinitely many 
decompositions of a density matrix $\rho$, and the first solutions to 
the convex roof of the threetangle emerged therefore only for 
special symmetric states\cite{LOSU,KENNLINIE,Jung09,HigherRankTau3}.
A considerable advance was the lower bound of the threetangle in terms of that of 
the $GHZ$ symmetrized version of the state. The convex roof of the square root of the 
threetangle\cite{VerstraeteDMV02,ViehmannII} of the latter 
could be calculated exactly\cite{Eltschka2012,Siewert2012} 
in the same way as outlined in Ref.~\cite{LOSU,KENNLINIE}.
Recently, a code was invented that upper bounds arbitrary $SL$ invariant entanglement measure $E$ 
for density matrices of variable rank\cite{Rodriquez14}, 
by considering the manifold made of the extreme points of the 
zero-polytopes\footnote{A zero-polytope is made out of those pure states on which $E$ vanishes.
It has been called {\em zero-simplex} in refs.~\cite{LOSU,KENNLINIE}. Every such zero-polytope is made of
zero-simplices of $d+1$ points in $d$ dimensions.}, 
and measuring the distance to the barycenter of these points.
This method has been applied for making statements about
the residual tangle in pure states of four qubits\cite{AdessoRegula14}.
This work fills the gap, proclaimed in this latter reference, 
namely of providing with an atlas of the threetangle 
for the existing classification of four-qubit pure states\cite{VerstraeteDMV02}. 

This article is laid out as follows. In the following section, I review relevant subjects of
how convex roofs can be obtained and the connection 
to what is called {\em characteristic curves}\cite{LOSU,KENNLINIE}. 
In the sequel, I will 
treat each different class of $SL(2)^{\otimes 4}$ separately and calculate the convex roof, except for one class, 
where an upper bound is obtained. At the end, I make some concluding remarks.

\section{Minimal characteristic curve and convex roof}

\begin{figure}
  \includegraphics[width=\linewidth]{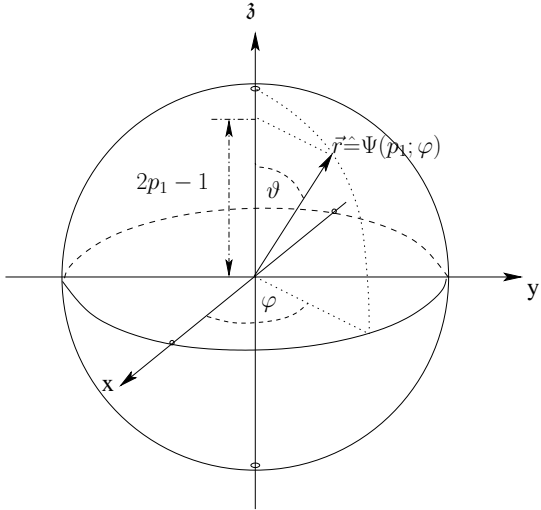}
  \caption{The Bloch sphere with radius $|\vec{r}|=1$ and the standard parametrization of its surface
in the angles $\vartheta$ and $\varphi$ of the vector $\vec{r}$. This vector corresponds to
the pure state $\ket{\Psi(p_1;\varphi)}=\cos{\frac{\vartheta(p_1)}{2}}\ket{\psi_{1}}+ \sin{\frac{\vartheta(p_1)}{2}}e^{i\varphi} \ket{\psi_{2}}=\sqrt{p_1}\ket{\psi_{1}}+ \sqrt{1-p_1}e^{i \varphi}\ket{\psi_{2}}$. The $r_{\frak{z}}$ component is related to the probability $p_1$ via $r_{\frak{z}}=2p_1-1$.}
  \label{blochsphere}
\end{figure}

Let $E$ denote the entanglement measure, and let it be a homogenous polynomial $SL$ invariant
of degree $D=2n$ for integers $n$.
The entanglement of a rank two density matrix 
\beq
\rho=\sum_{i=1}^2 p_i \ketbra{\psi_i}\; ;\ 
\braket{\psi_i}{\psi_j}=\delta_{ij}
\eeq
can only be made out of pure states in its range 
\beq
\ket{\Psi(p_1;\varphi)}=\sqrt{p_1}\ket{\psi_1}+\sqrt{p_2}e^{i\varphi} \ket{\psi_2}\; .
\eeq
These states are best visualized on a Bloch sphere (see Fig. \ref{blochsphere}).
In order to calculate the convex roof, 
we have to analyze the entanglement $E(\ket{\Psi(p_1;\varphi)})$ for all such states.
This corresponds to looking at\cite{LOSU,KENNLINIE}
\beq
E(\ket{\Psi(z)})=E(\ket{\psi_1}+z \ket{\psi_2})\;,
\eeq
where the $p_i$ are functions of $z$: $p_1=1/(1+|z|^2)$ and $p_2=|z|^2/(1+|z|^2)$.
The characteritic curves for $E$ are $E(\ket{\Psi(p_1;\varphi)}=E(\ket{\Psi(z)}$ and their 
minimum (which may depend explicitly on $\varphi$)
\beq
E_{\min}(p):=\min_{\varphi}\{E(\ket{\Psi(p;\varphi)} \}
\eeq
are of utmost importance for the determination of $E(\rho)$\cite{LOSU,KENNLINIE}.
I have relaxed the definition of the characteristic curve of Ref.~\cite{LOSU,KENNLINIE} 
slightly: the minimal characteristic curve $E_{min}(p)$ here is the characteristic curve in
Ref.~\cite{LOSU,KENNLINIE}.

Whereas in general one will find an explicit phase dependence $p(\varphi)$ in $E_{\min}$, 
here we have given values for $\varphi=\varphi_0$ for which the minimum is achieved. 
If a decomposition is found whose entanglement lies on the convexified minimal characteristic
curve, then this decomposition is certainly optimal\cite{LOSU,KENNLINIE} and the convex roof
is obtained. 
In general it is a solution which lies above the convexification of $E_{\min}$.

We need to consider solutions to the zero-polytope made out of the solutions to
\beq
E(\frac{\ket{\psi_1}}{|\ket{\psi_1}|}+z \frac{\ket{\psi_2}}{|\ket{\psi_2}|})=0
\eeq
and define 
\beq\label{charcurves}
\ket{\Psi_{z}}:= \sqrt{p_1(z)} (\frac{\ket{\psi_1}}{|\ket{\psi_1}|}+z\frac{\ket{\psi_2}}{|\ket{\psi_2}|})
\eeq
and the probability $p_1(z)$ of $\ket{\psi_1}$ is
\beq
p_1(z)=\frac{1}{1+|z|^2}\; .
\eeq
The zero-polytope is made out of as many pure states as the homogeneous degree of $E$.
For $E=\tau_3$ the homogeneous degree is $4$ and the (convex) zero-polytope becomes a zero-simplex.

Henceforth, I will consider pure states of four qubits, whose three qubit minors
automatically have rank two, and I will consider $\tau_3$ and $\sqrt{\tau_3}$ as entanglement
measures. The threetangle $\tau_3$ is defined as\cite{Coffman00} 
(see also in refs.~\cite{Wong00,VerstraeteDM03,OS04})
\nbeqa
\tau_3 &=& |d_1 - 2d_2 + 4d_3|  \\
  d_1&=& \psi^2_{000}\psi^2_{111} + \psi^2_{001}\psi^2_{110} + \psi^2_{010}\psi^2_{101}+ \psi^2_{100}\psi^2_{011} \\
  d_2&=& \psi_{000}\psi_{111}\psi_{011}\psi_{100} + \psi_{000}\psi_{111}\psi_{101}\psi_{010}\\ 
    &&+ \psi_{000}\psi_{111}\psi_{110}\psi_{001} + \psi_{011}\psi_{100}\psi_{101}\psi_{010}\\
    &&+ \psi_{011}\psi_{100}\psi_{110}\psi_{001} + \psi_{101}\psi_{010}\psi_{110}\psi_{001}\\
  d_3&=& \psi_{000}\psi_{110}\psi_{101}\psi_{011} + \psi_{111}\psi_{001}\psi_{010}\psi_{100}\ \ ,
\neeqa

In order to avoid misunderstandings, 
the function is highlighted whose convex roof will be calculated by a hat symbolizing the convex roof. 
It is worth mentioning here that the 
convex roofs $\widehat{\tau}_3$ of $\tau_3$ and $\widehat{\sqrt{\tau_3}}$ of $\sqrt{\tau_3}$ are 
different in general; in particular we have $\widehat{\sqrt{\tau_3}}^2\leq\widehat{\tau}_3$.
In fact, for every positive invertible concave function $f$ we have 
$f^{-1}(\widehat{f(\tau)})\leq \widehat{\tau}$. This is because for an optimal decomposition of $\tau$,
we get with $f(\tau)$ a result, which in some points may not be convex any more. 
Since for calculating the convex roof, one has to convexify $f(\tau)$, and
any optimal decomposition of $f(\tau)$ will lie below this result, we finally get
$f^{-1}(\widehat{f(\tau)})\leq\widehat{\tau}$. \\
I will consider the functions $\widehat{\sqrt{\tau_3}}^2$ and $\widehat{\tau}_3$ in what follows.
I will sometimes call both functions {\em threetangle}.

\section{Threetangle of pure four-qubit states}\label{central}

A classification of four-qubit states has been found in Ref.~\cite{VerstraeteDMV02}, 
which can be compared to the Schmidt decomposition for two qubits
and its extension for three qubits\cite{Acin00}, where however the local transformations
are out of the general linear group with non-vanishing determinant, hence not unitary in general.

Here, I present calculations that offer in general an upper bound to the threetangle.
In most cases however, they provide with the exact convex roof of the threetangle for the states.
I looked at all nine classes whereas discussions are only included up to class 6, 
since for the remainig classes 7 to 9 it is either clear that the respective threetangle vanishes, 
or the previously given estimate coincides with the convex roof.
A good upper bound is given for class 5, 
where I have good reasons why the decomposition that I obtain is at least close to optimal.

The qubits are numbered as
$(1,2,3,4)$ corresponding to the state $\ket{q_1,q_2,q_3,q_4}$. 
Please notice that the two states $\ket{\psi_i}$, $i=1,2$, are not normalized. 
All subsequent definitions are class-local.

\subsection{Class 1}
The representant of this class, which contains all the stochastic states\cite{VerstraeteDMV02,VerstraeteDM03,OS04,OS05}
with reduced local density matrices proportional to the identity, is
\beqa
\ket{G_{abcd}^1}&=&\frac{a+d}{2}(\ket{0000}+\ket{1111}) \nonumber\\
&& + \frac{a-d}{2}(\ket{0011}+\ket{1100})  \nonumber\\
&& +\frac{b+c}{2}(\ket{0101}+\ket{1010})  \nonumber\\
&&+ \frac{b-c}{2}(\ket{0110}+\ket{1001})\label{class1}
\eeqa
All three qubit minors are equivalent with respect to their eigenvalues and 
threetangle in their eigenstates,
and we have $\tau_3[\rho]=0$, as pointed out already in Ref.~\cite{VerstraeteDMV02}. 

\subsection{Class 2}

\begin{figure}
\centering
  \includegraphics[width=.9\linewidth]{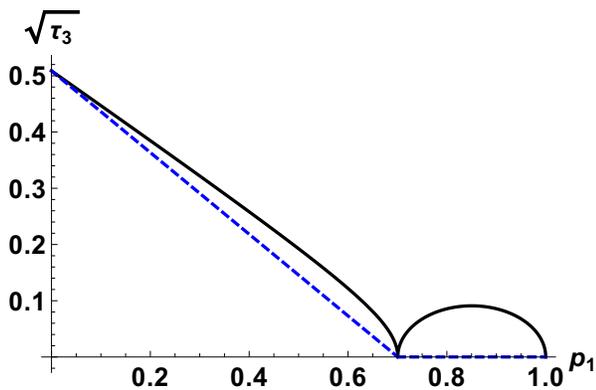}
  \caption{The minimal characteristic curves for the angles $\varphi=\arg{z_{0}}+j\pi$ 
on the Bloch sphere are shown for $j=0,1$ (black line) together with the convexified 
curve $\tau_3^{conv}$ (blue dashed line). The optimal decompositions are $\ket{\psi_{2}}$ and the 
two corresponding $\ket{\psi_{z_{0}}}$ for a finite value of 
$\sqrt{\tau_3}$, and $\ket{\psi_{1}}$ and the two $\ket{\Psi_{z_{0}}}$ for vanishing threetangle 
(see explanations in the text).}
  \label{sqrttau3:2}
\end{figure}
The representant for the second class is defined as\cite{VerstraeteDMV02}
\beqa
\ket{G_{abc}^2}&=&\frac{a+b}{2}(\ket{0000}+\ket{1111}) \nonumber\\
&&+ \frac{a-b}{2}(\ket{0011}+\ket{1100})\label{class2}\\
&&+c(\ket{0101}+\ket{1010}) + \ket{0110}\; ,\nonumber
\eeqa
and the reduced three-qubit density matrices are again equivalent, in that they have the same eigenvalues
and also the same threetangles of their eigenstates.
Tracing out e.g. the fourth qubit, this gives rise to the two eigenstates 
\beqa
\ket{\psi_1}&=& (a^*-b^*)\ket{001}+(a^*+b^*)\ket{111}\nonumber\\
&&+2c^*\ket{010}\\
\ket{\psi_2}&=& (a^*+b^*)\ket{000}+(a^*-b^*)\ket{110}\nonumber\\
&&+2c^*\ket{101}+2\ket{011}\; ,
\eeqa
with the probability of $\ket{\psi_1}$ being
\beq
p_{1}(a,b,c)=\frac{|a|^2+|b|^2+2|c|^2}{2 (1+|a|^2+|b|^2+2|c|^2)}\; .
\eeq

\begin{figure}
\centering
  \includegraphics[width=.9\linewidth]{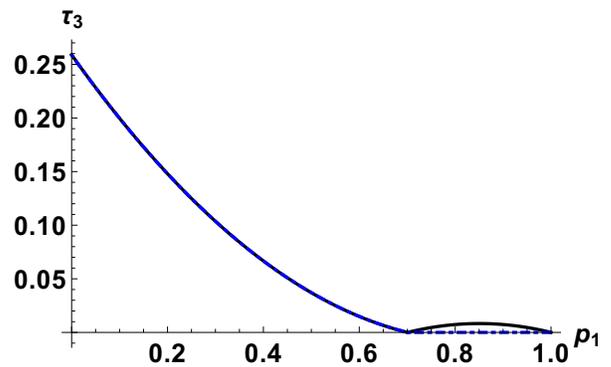}
  \caption{A typical example for a minimal characteristic curve for $\tau_3$ is shown (black line)
together with the convexified 
one $\tau_3^{conv}$ (blue dashed line). Here the curve is strictly convex where the threetangle is finite.
This leads to an optimal decomposition at $p_1$ which is $\ket{\Psi_{\pm z(p_1)}}$ and gives rise to  
smaller values for the convex roof $\widehat{\tau}_3$.}
  \label{tau3:2}
\end{figure}

A typical example of a minimal characteristic curve is shown in figs.~\ref{sqrttau3:2} and \ref{tau3:2}
for $\sqrt{\tau_3}$ and $\tau_3$, respectively.
The convexified minimum of the characteristic curves (blue dashed line), $\sqrt{\tau_3^{conv}}(p_1)$
and $\tau_3^{conv}(p_1)$, are also plotted. A decomposition of a convexified region is always given by the 
corresponding states $\ket{\Psi_{z_0}}$, which flank the region. 
The zero simplex is flanked by two trivial zeros at $p_1=1$ ($z=0$) and the two values
\beqa
z_{0}&=&\pm\frac{1}{2}\sqrt{\frac{2+|a-b|^2+2|c|^2}{|a|^2+|b|^2+2|c|^2}}\nonumber\\
&&\times \sqrt{\frac{b^2}{c}\left(1+\frac{c^2}{b^2}\frac{a(a-b)+c^2}{(a-b)^2}\right)}^*\; ,
\eeqa
and hence $p_{0}=1/(1+|z_{0}|^2)$.
Whereas for $\tau_3$ the only convexified region is inside the zero-simplex\cite{LOSU,KENNLINIE}
in the interval $[p_{0},1]$,
for $\sqrt{\tau_3}$ it is given additionally by a straight line connecting the threetangle at $p=0$ 
with the beginning of the zero-simplex at $p_{0}$.
Hence, the convex roof $\widehat{\sqrt{\tau_3}}$ of $\sqrt{\tau_3}$ is
\begin{widetext}
\beq
\widehat{\sqrt{\tau_3}}=\max\left\{0,2\sqrt{\frac{\left|(a^2-b^2)c\right|}{1+|a|^2+|b|^2+2|c|^2}}
\left(1-\frac{|a|^2+|b|^2+2|c|^2}{2 p_{0;2}(1+|a|^2+|b|^2+2|c|^2)}\right)\right \}\; ,
\eeq
\end{widetext}
and the optimal decomposition in between $p_1=0$ and $p_{0}$ is made of the eigenstate at $\ket{\psi_{2}}$ 
and the two states at $\ket{\Psi_{\pm z_{0}}}$. 
For $\tau_3$ instead, we obtain that the minimal characteristic curve is convex in $[p_1=0,p_{0}]$, 
and hence the optimal decomposition 
is given here in terms of $\ket{\Psi_{z(p)}}$ and $\ket{\Psi_{-z(p)}}$ between $p_1=0$ and $p_{0}$.
\footnote{Here, with $z(p_1)$ I mean that $\arg z=\arg z_0$ of the solutions $z_0$ of $E(z)=0$  ($E=\tau_3$ here), 
since here the minimal characteristic curve does not depend on $\arg z=\varphi$.}
Whereas we managed to give a closed formula for the convex roof $\widehat{\sqrt{\tau_3}}$,
we here give $\widehat{\tau}_3$ only in the implicit form
\beq
\widehat{\tau}_3=\max\{0,\tau_3^{conv}(p_{1}(a,b,c))\}\; ,
\eeq
with the convexification $\tau_3^{conv}(p_{1})$ of $\tau_3(p_{1})$.
\begin{figure}
\centering
  \includegraphics[width=.9\linewidth]{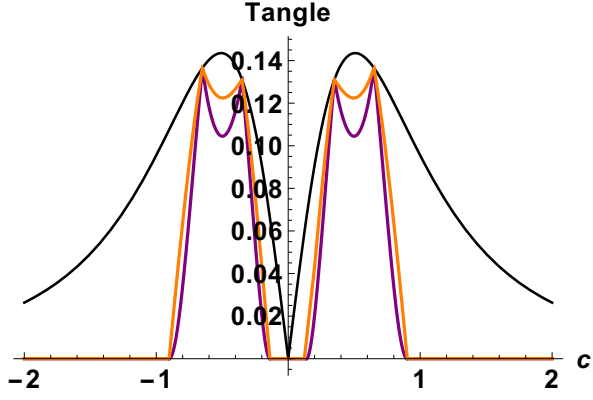}
  \caption{Class 2: The convex roofs are shown for $\widehat{\tau}_3$ (central orange curve) and
$\widehat{\sqrt{\tau_3}}^2$ (lower purple curve) for real values of $a=0.65$, $b=0.35$ and $c$. It is seen that 
$\widehat{\sqrt{\tau_3}}^2<\widehat{\tau}_3$. For comparison we plot also the curve obtained
by Ref.~\cite{AdessoRegula14} from the algorithm suggested in Ref.~\cite{Rodriquez14} (upper black curve).
They coincide at the pikes of the curves, where the convex roof is given by the 
eigen-decomposition.}
  \label{tau3s.65.35:2}
\end{figure}

To illustrate the convex roofs, we show $\widehat{\sqrt{\tau_3}}^2$ and
$\widehat{\tau}_3$ for $a=0.65$, $b=0.35$ in fig.~\ref{tau3s.65.35:2} 
as a function of a real $c$. The two convex roofs are seen to
rise sharply at $c=\pm 0.1388$ and vanish again at $c=\pm 0.9022$ 
with two cusps at $c=\pm 0.35$ with values $\widehat{\tau}_3=0.1311$, 
and at $c=0.65$ with value $\widehat{\tau}_3=0.1366$.
Here, the characteristic curves become straight lines for
$\sqrt{\tau_3}$ and the corresponding parabolas for $\tau_3$. Hence, both convex roofs
coincide here.
For comparison we show the curve from~\cite{AdessoRegula14,AdessoRegula14b}
\beq\label{AdessoRegula}
\widehat{\sqrt{\tau_3}}^2=4 \frac{|(a^2-b^2)c|}{(1+|a|^2+|b|^2+2|c|^2)^2}
\eeq
which is an upper bound to $\widehat{\sqrt{\tau_3}}^2$, as it should be. 
It clearly overestimates the threetangle in the state
except at the cusps at $c=\pm a$ and $c=\pm b$, where all zeros $z_0$ coincide.

\subsection{Class 3}
\begin{figure}
\centering
  \includegraphics[width=.9\linewidth]{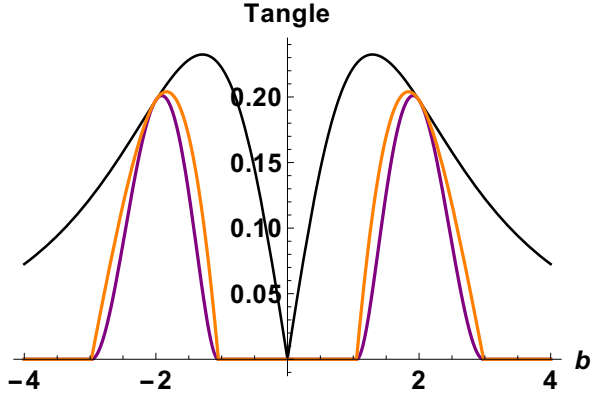}
  \caption{Class 3: The convex roofs $\widehat{\tau_3}$ (central orange curve) and 
$\widehat{\sqrt{\tau_3}}^2$ (lower purple curve) are shown together 
with the estimate of Ref.~\cite{AdessoRegula14} (upper black curve) for $a=2.0$. 
It can be seen that they both coincide at $a=b=2.0$, where the minimal
characteristic curve of $\widehat{\sqrt{\tau_3}}$ is a straight line. 
At this point, it also must coincide with the estimate in Ref.~\cite{AdessoRegula14}, 
which highly overestimates the convex roofs. The optimal decompositions are
made out of $\ket{\Psi_{\pm z_{0}}}$ and $\ket{\psi_2}$ for a finite $\widehat{\sqrt{\tau_3}}$, 
and $\ket{\Psi_{\pm z(p_1)}}$ for finite $\widehat{\tau}_3$.}
  \label{tau3s.2:3}
\end{figure}

The representant of the third class is given by\cite{VerstraeteDMV02}
\beqa
\ket{G^3_{ab}}&=&a (\ket{0000}+\ket{1111})\nonumber\\
&& + b(\ket{0101}+\ket{1010}) \nonumber\\
&&+\ket{0110}+\ket{0011}\label{class3}
\eeqa
There are two inequivalent three-qubit density matrices with respect to 
their eigenvalues and threetangle here. The threetangle vanishes for two of these 
density matrices (qubits numbers 2 and 4 traced out). 
The eigenstates of the remaining reduced density matrices (here: qubit 1 traced out) are
\beqa
\ket{\psi_1}&=&b^*\ket{010}+a^*\ket{111}\\
\ket{\psi_2}&=&a^*\ket{000}+b^*\ket{101}+\ket{011}+\ket{110}
\eeqa
and the typical characteristic curves have mainly the same form as for class 2, and are 
therefore not shown.
The weight of $\ket{\psi_1}$ in the density matrix is
\beq
p_{1}(a,b)=\frac{|a|^2+|b|^2}{2(1+|a|^2+|b|^2)}\; .
\eeq
We consider again solutions to
\beq
\tau_3(\frac{\ket{\psi_1}}{|\ket{\psi_1}|}+z \frac{\ket{\psi_2}}{|\ket{\psi_2}|})=0
\eeq
which has zeros at
\beq
z_{0}=\pm i\left(\frac{a^2-b^2}{2\sqrt{ab}}\right)^*\sqrt{\frac{2+|a|^2+|b|^2}{|a|^2+|b|^2}}
\eeq
and corresponding values for $p_{0}$.
This leads to the following convex roof of $\sqrt{\tau_3}$
\beq
\widehat{\sqrt{\tau_3}}=\max\left\{0, \frac{4|ab|-\left |a^2-b^2\right|^2}{2\sqrt{|ab|}(1+|a|^2+|b|^2)} \right\}\; .
\eeq
The corresponding threetangles $\widehat{\sqrt{\tau_3}}^2$ and
$\widehat{\tau}_3$ are shown for $a=2.0$ in fig.~\ref{tau3s.2:3}, where it is seen that 
$\widehat{\sqrt{\tau_3}}^2<\widehat{\tau}_3$ except for the points 
$b\approx \pm 1.0498$ and $b\approx \pm 2.9804$, where the threetangle vanishes, and at
$b=\pm 2.0$, where the charakteristic curve of $\sqrt{\tau_3}$ again reduces to a straight line 
and the zeros $z_{0}$ coincide.
At this latter points the convex roof again coincides with the upper bound of~\cite{AdessoRegula14}, 
since the
method of ref.~\cite{Rodriquez14} gives the exact convex roof here. It is seen however that in general the
estimate taken from~\cite{AdessoRegula14} considerably overestimates the threetangle of the state.

\subsection{Class 4}
\begin{figure}
\centering
  \includegraphics[width=.9\linewidth]{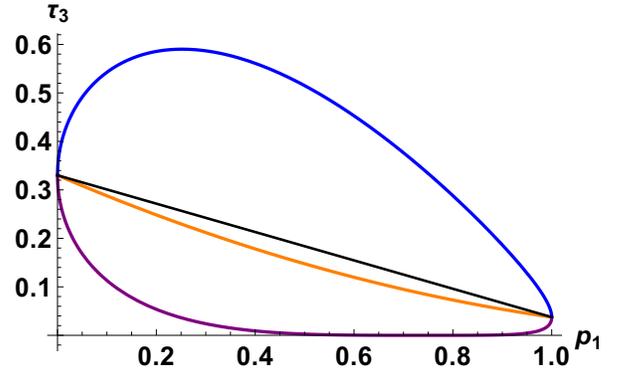}
  \caption{Some characteristic curves are shown for real values $a=0.4$, $b=1.9$ and 
angles $\varphi=\pi$ (the minimum; lower purple curve), 
$\varphi=\pm\pi/2$ (central orange curve), and $\varphi=0$ (the maximum; upper blue curve). 
For comparison the straight line (thin black curve) marks the convex combination of the two eigenstates. 
There is a single zero of fourth order at
$z=-|\ket{\psi_2}|/|\ket{\psi_1}|$ of the equation 
$\tau_3(\ket{\psi_1}/|\ket{\psi_1}|+z \ket{\psi_2}/|\ket{\psi_2}|)=0$.}
  \label{tau3:4}
\end{figure}

\begin{figure}
\centering
  \includegraphics[width=.9\linewidth]{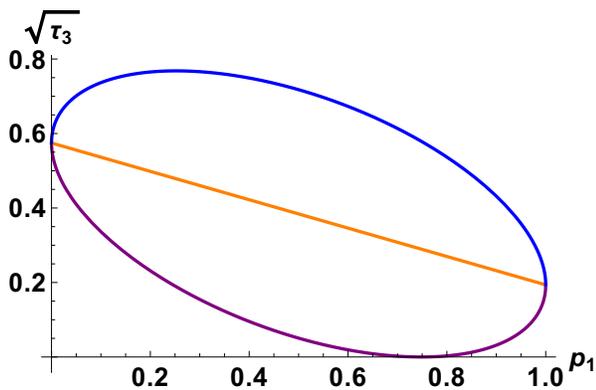}
  \caption{The minimal (lower purple curve) and maximal (upper blue curve) characteristic curves 
are shown here for real $a=0.4$ and $b=1.9$ and 
the same angles as in fig.~\ref{tau3:4}. 
The curves at $\varphi=\pm \pi/2$ coincide and possibly even give the convex roof result.
Here, this is a straight line connecting both eigenstates, and consequently also the algorithm of
Ref.~\cite{Rodriquez14} gives this result.}
  \label{sqrttau3:4}
\end{figure}

The class 4 is represented by\cite{VerstraeteDMV02}
\beqa
\ket{G^4_{ab}}&=&a (\ket{0000}+\ket{1111}) + \frac{a+b}{2}(\ket{0101}+\ket{1010}) \nonumber\\
&+&\frac{a-b}{2}(\ket{0110}+\ket{1001})\nonumber \\
&+& \frac{i}{\sqrt{2}}(\ket{0001}+\ket{0010}+\ket{0111}+\ket{1011})\label{Class4}
\eeqa
Also here, all four reduced density matrices are equivalent with respect to their eigenvalues and 
threetangle of their eigenstates.
A set of eigenstates of one of the reduced three-qubit density matrix (qubit number 1 traced out) is
\beqa
\ket{\psi_1}&=&i\sqrt{2} a^*s_-\ket{000}+(s_- -2a(a^*-b^*))\ket{001}\nonumber\\
&+& (s_- -2a(a^*+b^*))\ket{010}+2 i\sqrt{2} a\ket{011}\nonumber\\
&+& \frac{i}{\sqrt{2}}s_-(a^*+b^*)\ket{101}\\
&+&\frac{i}{\sqrt{2}}s_-(a^*-b^*)\ket{110} +(s_- -4|a|^2)\ket{111}\nonumber\\
\ket{\psi_2}&=&i\sqrt{2} a^*s_+\ket{000}+(s_+ +2a(a^*-b^*))\ket{001}\nonumber\\
&+& (s_+ +2a(a^*+b^*))\ket{010} -2 i\sqrt{2} a\ket{011}\nonumber\\
&+& \frac{i}{\sqrt{2}}s_+(a^*+b^*)\ket{101}\\
&+&\frac{i}{\sqrt{2}}s_+(a^*-b^*)\ket{110} +(s_+ +4|a|^2)\ket{111}\; ,\nonumber
\eeqa
where $s_\pm=\sqrt{1+8|a|^2}\pm 1$,
with a typical set of characteristic curves shown in fig.~\ref{tau3:4} and~\ref{sqrttau3:4} 
for $\tau_3$ and $\sqrt{\tau_3}$, respectively. 
The weight of $\ket{\psi_1}$ is
\beq
p_{1}(a,b)=\frac{2+3|a|^2+|b|^2-\sqrt{1+8|a|^2}}{4+6|a|^2+2|b|^2}\; .
\eeq
We show the minimal curve in purple (at $\varphi=\pi$), the two identical central curves in orange ($\varphi=\pm \pi/2$),
and the maximum ($\varphi=0$, blue curve). 
There is a four-fold degeneracy of the solution to the equation
$\tau_3(\ket{\psi_1}/|\ket{\psi_1}|+z \ket{\psi_2}/|\ket{\psi_2}|)=0$, 
which is $z_0=-|\ket{\psi_2}|/|\ket{\psi_1}|$;
it corresponds to $p_{0}=|\ket{\psi_1}|/(|\ket{\psi_1}|+|\ket{\psi_2}|)$.\\
It is worth mentioning that the minimal characteristic curve can impossibly constitute the minimum of 
decompositions of $\rho$, because it had to be combined with curves beyond the $\varphi=\pm \pi/2$ 
lines to this end. 
One possibility is to combine it with the curve at $\varphi=0$ in order to give a decomposition of $\rho$, 
which is the maximal curve, though.
In this case the result for $\sqrt{\tau_3}$ (fig.~\ref{sqrttau3:4})
is the same as the central orange curves at $\varphi=\pm \pi/2$.
This does not change, when whatever $q_i\in [0,1]$ is taken as defining 
the decomposition at an arbitrary angle. In fact, in the case of coinciding zeros $z_0$ 
it has been shown during the publication process of this work\cite{AdessoRegula16} 
that the exact convex roof is obtained for $\widehat{\sqrt{\tau_3}}$, and that all decompositions of $\rho$ 
are optimal.

Therefore we conclude that $\widehat{\sqrt{\tau_3}}^2$ as well as 
$\widehat{\tau}_3$ is given by
\beq
\widehat{\sqrt{\tau_3}}^2=\widehat{\tau}_3=\max\{0,2\frac{\left | a^2-b^2 \right |}{2+3|a|^2 + |b|^2}\}\; .
\eeq

We want to mention however that
the real outcome of the analysis of Ref.~\cite{AdessoRegula14} is that consequently the residual tangle is larger 
than a negative value, due to violations in this class 4 only. 
In ref.~\cite{AdessoRegula14} instead, it was not known that it was the convex roof for that class, and
the authors could at best say that an extended monogamy clearly guarantees positivity of the residual tangle.
This changes with the Ref.~\cite{AdessoRegula16} in which it is clearly demonstrated to be 
the convex roof.
\begin{figure}
\centering
  \includegraphics[width=.9\linewidth]{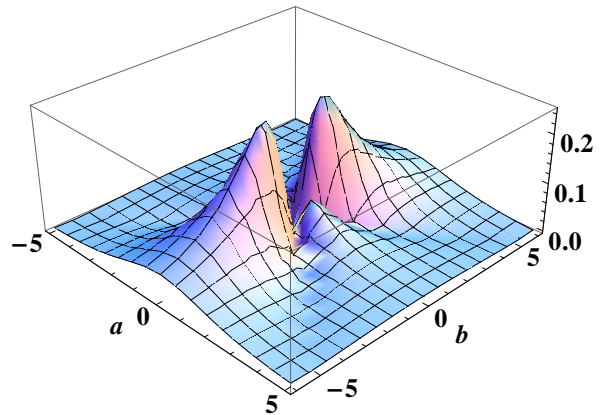}
  \caption{The values for $\widehat{\tau_3}=\widehat{\sqrt{\tau_3}}^2$ in class 4 
for real values of $a$ and $b$.}
  \label{tangles:4}
\end{figure}

We show the values for $\widehat{\tau}_3$ and $\widehat{\sqrt{\tau_3}}^2$ for real values of $a$ and $b$ in fig.~\ref{tangles:4}.

\subsection{Class 5}

The representant of the fifth class is given by\cite{VerstraeteDMV02}
\beqa
\ket{G^5_{a}}&=&a (\ket{0000}+\ket{1111}+\ket{0101}+\ket{1010}) \nonumber\\
&&+ i(\ket{0001}-\ket{1011})+\ket{0110}\label{Class5}
\eeqa
The reduced three-qubit density matrices are all equivalent with respect to the eigenvalues.
But there are two inequivalent subsets when looking at the threetangles of their eigenstates. 
The corresponding set of eigenstates of one of the reduced three-qubit density matrices is
\beqa
\ket{\psi_{1,1}}&=&-i\ket{000}+a^*\ket{010}+i\ket{101}+ a^*\ket{111}\\
\ket{\psi_{1,2}}&=&a^*\ket{000}+\ket{011}+ a^*\ket{101}
\eeqa
when qubit number 4 (this is equivalent to tracing out qubit number 2) is traced out, and
\beqa
\ket{\psi_{2,1}}&=&\ket{010}+a^*\ket{100}+i\ket{101}+ a^*\ket{111}\\
\ket{\psi_{2,2}}&=&a^*\ket{000}-i\ket{001}+ a^*\ket{011}
\eeqa
when tracing out qubit number 3.
The weight or probability of the states $\ket{\psi_{1,1}}$ and $\ket{\psi_{2,1}}$ is
\beq
p_{1}(a)=\frac{2(1+|a|^2)}{3+4|a|^2}\; .
\eeq

We will consider the former case first.
The solution for the zero-simplex is a four-fold zero at $z=0$ here corresponding to
a solution for the convex roof of
\beq
\widehat{\tau}_3=\widehat{\sqrt{\tau_3}}^2=16\frac{|a|^2}{(3+4|a|^2)^2}\; ,
\eeq
which naturally is the same as the estimate of Ref.~\cite{AdessoRegula14}.

\begin{figure}
\centering
  \includegraphics[width=.9\linewidth]{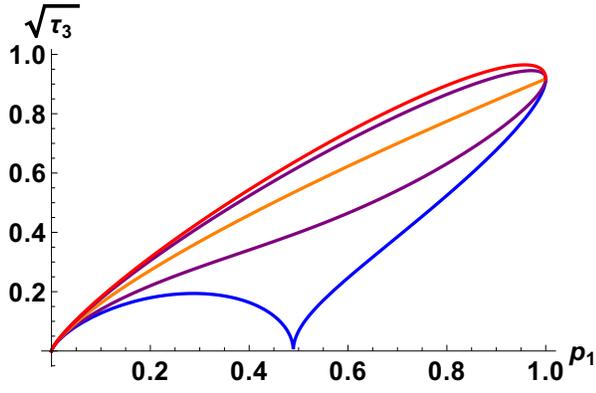}
  \caption{For real value $a=0.3$ among some characteristic curves are shown in particular 
the minimal characteristic curve at $\varphi=\pi/2$ (lower blue line)
and maximal characteristic curve at $\varphi=3\pi/2$ (higher red curve) and the two
intermediate coinciding curves at $\varphi=0,\pi$ (central orange curve).}
  \label{sqrttau3:5.2}
\end{figure}

\begin{figure}
\centering
  \includegraphics[width=.9\linewidth]{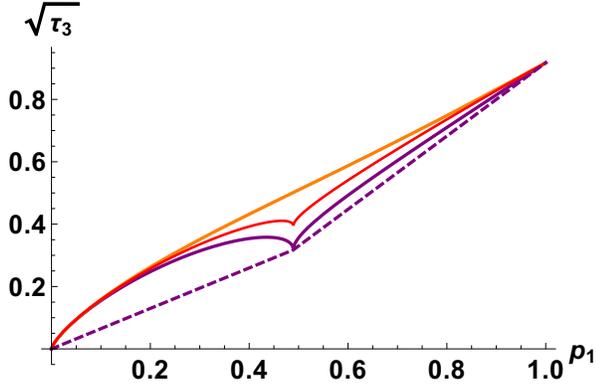}
  \caption{For $a=0.3$ the sum of two by an angle of $\pi$ differing characteristic curves are shown for $\varphi=0,\pi$ 
(higher orange curve) and for $\varphi=\pi/2,3\pi/2$ (lower purple curve), combining to possible decompositions of $\rho$.
The convexification of this latter curve is also shown (purple dashed line).
A decomposition into three states (thin red curve) is shown to be situated between the lower purple and higher orange curve.}
  \label{sqrttau3-combi:5.2}
\end{figure}

The latter case is more interesting, and is given by the characteristic curves, 
which I show in fig.~\ref{sqrttau3:5.2}.
Here is a threefold degenerate solution for the zero-simplex at $z=0$ and a 
single solution at $z_{0}=8\sqrt{2}i(a^*)^2\sqrt{\frac{1+|a|^2}{1+2|a|^2}}$
to $\tau_3(z\ket{\psi_{2,1}}/|\ket{\psi_{2,1}}|+\ket{\psi_{2,2}}/|\ket{\psi_{2,2}}|)=0$, 
which corresponds to the probability
\beqa
p_{0}&=&1-\frac{1}{1+|z_{0}|^2}\nonumber\\
&=&\frac{128|a|^4(1+|a|^2)}{1+2|a|^2+128|a|^4(1+|a|^2)}\; .
\eeqa
The results for certain decompositions of $\rho$ are shown in fig.~\ref{sqrttau3-combi:5.2}. 
At first, I am considering the decompositions out of the two states
at angles $\varphi=\pm \pi/2$ (lowest purple curve) and at $\varphi=0,\pi$ (highest orange curve)
at a given $p_1$. 
The combination of three states with phases $\varphi=3i\pi/2+2m i\pi/3$ for $m=0,1,2$
(red curve) is shown to lie in between the purple and the orange curve.

\begin{figure}
\centering
  \includegraphics[width=.9\linewidth]{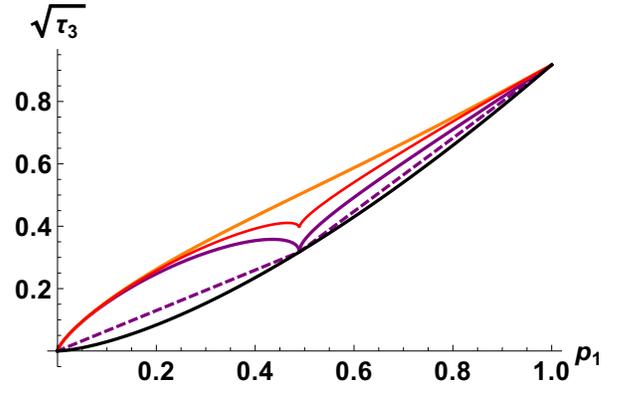}
  \caption{In addition to figure \ref{sqrttau3-combi:5.2}, 
convex combinations of the state $\ket{\Psi_{z_{0}}}$ with a second state at $\varphi=\pi/2$ 
to give a decomposition of $\rho(p_1)$ are shown in the convex lower black curve. Every combination from the zero-simplex 
with another state at $\varphi=\pi/2$ lies above this curve. This is the lowest upper bound.
I conjecture that it equals the convex roof.}
  \label{sqrttau3-combiplus:5.2}
\end{figure}

Another choice is, to combine the state at $p_{0}$ at $\varphi = \pi/2$ with another state at the angle 
$\varphi=3\pi/2$ and $q$. Any two such states form a decomposition for 
$\rho(p_1)=p_1\ketbra{\psi_{2,1}}+(1-p_1)\ketbra{\psi_{2,2}}$ with some $p_1(q)$, 
giving a value for $\sqrt{\tau_3}(p_1)$
which is shown in fig.~\ref{sqrttau3-combiplus:5.2}. This curve is convex 
(lowest black curve in fig.~\ref{sqrttau3-combiplus:5.2})
and always below the preceding convexified minimal curve (dashed purple curve), 
except at $p_{0}$ and at $p_1=0,1$, where it coincides with it. 
\footnote{Note that this must not be confused with the convex characteristic curve,
which of course lies below this curve.}
Each curve corresponding to a sum 
of some pure state at $q_0$ on the minimal characteritic curve at $\varphi=\pi/2$
and the to $\rho(p_1)$ corresponding pure state at $3\pi/2$ is lying above this curve
and this does not change if choosing some other angle $\varphi$.
It is futhermore the minimal curve 
of superpositions of a (in general mixed) state inside the zero-simplex at $\varphi=\pi/2$ and a corresponding pure state to $\rho(p_1)$
at $\varphi=3\pi/2$. 
It will therefore constitute a reasonable upper bound. I conjecture that it coincides 
with the convex roof.

\begin{figure}
\centering
  \includegraphics[width=.9\linewidth]{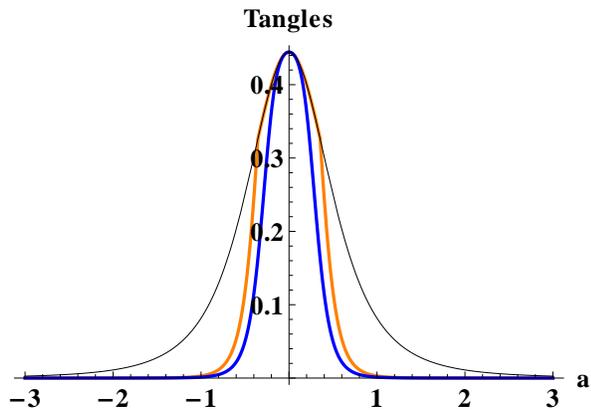}
  \caption{Class 5 for qubits 1 or 3 traced out: 
Upper bounds for $\widehat{\sqrt{\tau_3}}^2$ (which is the conjectured convex roof; lower blue curve)
and $\widehat{\tau}_3$ (central orange curve) are shown in this figure. They are compared
with the result in Ref.~\cite{AdessoRegula14} (upper black curve) 
making use of the algorithm related to Ref.~\cite{Rodriquez14}.
It is seen, that it overestimates the value for $\widehat{\sqrt{\tau_3}}$. It astonishingly coincides for 
small values of $a$ with the upper bound for $\widehat{\tau}_3$, but then deviates from it.
This coincidence is due to not having calculated the upper bound for the convexification of
the corresponding minimal curve that I have found. 
It is nicely seen that the upper bounds satisfy the inequaltiy $\widehat{\sqrt{\tau_3}}^2<\widehat{\tau}_3$
for the convex roofs.}
  \label{sqrttau3-upbound:5.2}
\end{figure}

I have calculated for this upper bound of $\widehat{\sqrt{\tau_3}}$ the entanglement in the state $\ket{G^5(a)}$
to be
\beqa\label{new:5}
\widehat{\sqrt{\tau_3}}^2&\leq& 4\frac{1+64|a|^2}{\left((3+4|a|^2)(1+64|a|^4)\right)^2}\\
&<&\frac{4}{(3+4|a|^2)^2}\; ,\label{old:5}
\eeqa
which is plotted in fig.~\ref{sqrttau3-upbound:5.2}.
For $\tau_3$ I get the implicit version 
\beq
\widehat{\tau_3}\leq \min\{0,\tau_3^{conv}(p_{1}(a))\}\; .
\eeq
It can be seen that the new upper bound \eqref{new:5} is considerably better than that 
previously published in Ref.~\cite{AdessoRegula14}, and given by Eq.~\eqref{old:5}.

I want to mention that in calculating the curve for $\tau_3$ I did not consider the convexified version of 
the corresponding curve $\tau^{conv}_3$. This would give a smaller result for the states under coonsideration. 
So while I conjectured that for $\sqrt{\tau_3}$ I have the convex roof, this certainly isn't so for $\tau_3$.
But I find it curious that it coincides with the upper bound in Ref.~\cite{AdessoRegula14} for 
$\widehat{\sqrt{\tau_3}}^2$.

\subsection{Class 6}

\begin{figure}
\centering
  \includegraphics[width=.9\linewidth]{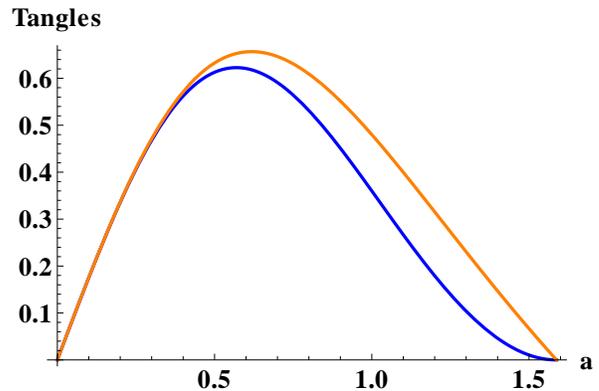}
  \caption{Class 6: Both convex roofs $\widehat{\sqrt{\tau_3}}^2$ (blue lower curve) and 
$\widehat{\tau}_3$ (orange curve) are shown. It is seen that 
$\widehat{\sqrt{\tau_3}}^2\leq\widehat{\tau}_3$. Both values vanish if $|a|\geq 2^{2/3}$.
It is strange that the upper bound from Ref.~\cite{AdessoRegula14} coincides with $\widehat{\sqrt{\tau_3}}^2$,
where instead something larger would be expected. This appears to be accidental and could be interrelated with
only two apparent solutions to the zero-simplex, when the $z$ appears in front of a state with zero threetangle.}
  \label{tangles:6}
\end{figure}

The representant of the fifth class is given by\cite{VerstraeteDMV02}
\beqa
\ket{G^6_{a}}&=&a (\ket{0000}+\ket{1111}) +\ket{0011}\nonumber\\
&&+\ket{0101}+\ket{0110}\label{Class6}
\eeqa
The reduced three qubit density matrix has non-vanishing threetangle only, when tracing out the first qubit;
it gives zero threetangle in the remaining cases. 
The typical characteristic curves for the non-vanishing case have the same structure as in class 2, 
and are hence not shown here.
The interesting eigenstates are
\beqa
\ket{\psi_{1}}&=&\ket{111}\\
\ket{\psi_{2}}&=&a^*\ket{000}+\ket{011}+ \ket{101} + \ket{110}\; ,
\eeqa
and the weight of $\ket{\psi_{1}}$ is $p_{1}(a)=|a|^2/(3+2|a|^2)$.  
The solutions for the zero-simplex are given by two trivial zeros $z=0$, and
\beq
z_{0}=\pm\frac{1}{2}\sqrt{a^*(3+|a|^2)}\; .
\eeq
The convex roofs are obtained in the same way as for class 2 before. 
For $\widehat{\sqrt{\tau_3}}^2$ we obtain
\beq
\widehat{\sqrt{\tau_3}}^2=\max\{0,\frac{|a|(|a|^3-4)^2}{(2|a|^2+3)^2}\}\; ,
\eeq
and for $\widehat{\tau}_3=\max\{0,\tau_3^{conv}(p_{1}(a))\}$ we get an implicit result.
The threetangles of the state are shown in fig.\ref{tangles:6}.
Surprisingly, here the upper bound obtained in Ref.~\cite{AdessoRegula14} coincides with the convex roof solution
$\widehat{\sqrt{\tau_3}}^2$ whereas the zeros indicate that the result should be actually bigger. 
This strange coincidence could have been caused by ignoring the two trivial zeros.

For the remaining classes, the estimates in Ref.~\cite{AdessoRegula14} from the algorithm in Ref.~\cite{Rodriquez14}
are precisely the convex roof measure for $\widehat{\sqrt{\tau_3}}^2$ which coincides with $\widehat{\tau}_3$ in 
these cases.

\section{Conclusions}\label{concls}
I have classified four-qubit pure states with respect to their mixed threetangle.
In order to do so, I have taken the representatives of the nine SL-classes from Ref.~\cite{VerstraeteDMV02}.
Those for general four-qubit states can be obtained by acting 
with $SL$ operations on the four qubits 
and considering that the threetangle is an $SL$ invariant\cite{ViehmannII}.
Whereas the convex roof could be obtained in 8 out of 9 classes in finding an optimal decomposition, 
for class $5$ better upper bounds than those existing in Ref.~\cite{AdessoRegula14} could be given. 
There are strong indicators that it is at least close to the convex roof. 
I compare the results with a method related to that published in 
Ref.~\cite{Rodriquez14} that has been used to obtain upper bounds 
in Ref.~\cite{AdessoRegula14,AdessoRegula14b} and found 
that the method gives the exact value for the threetangle in cases where the four solutions of the zero-simplex do
coincide. In these cases already the eigen-decomposition provides the convex roof.
Away from these points, the convex roof is considerably below this estimate.
For class 2, the upper bound of Ref.~\cite{AdessoRegula14} has been corrected\cite{AdessoRegula14b}
and fits nicely as an upper bound with my findings.
Finally, due to Ref.~\cite{AdessoRegula16} which shows that the upper bound 
obtained in Ref.~\cite{AdessoRegula14} coincides with the convex-roof, the strong monogamy 
strictly has to be applied (see also Ref.~\cite{AdessoOsterlohRegula16}). 

\section*{Acknowledgements}

During the completion of this work, I had intense and fruitful discussion with G. Adesso and B. Regula. 


\end{document}